\documentclass[final]{aipproc}
\newcommand\selectedlayoutstyle{8x11single}
\layoutstyle\selectedlayoutstyle
\SetInternalRegister\hbadness{8000} 
\usepackage{graphicx}   
\newdimen\captwidth   
\newdimen\figwidth   
\def\eg{{\it e.g. }}
\def\etal{{et al. }}  

\def\zu{\rm\,}     

\begin{document}

\title 
      [The Archeops balloon experiment]
      {Archeops: A balloon experiment to measure \\ CMB anisotropies with
        a broad range of angular sizes}

\classification{}
\keywords{TBD}

\author{A. Benoît}{
  address={Centre de Recherche sur les Tr\`es Basses Temp\'eratures,
25 Avenue des Martyrs BP166, F--38042 Grenoble Cedex 9, France},
}

\iftrue
\author{the Archeops Collaboration}{
  address={from the following institutes:\\
California Institut of Technology, Pasadena USA\\
Centre d'Etude Spatiale des Rayonnements, Toulouse France\\
Centre de spectrométrie nucléaire et de spectrométrie de masse, Orsay France\\
Colllège de France, Paris France\\
DAPNIA CEA, Saclay France\\
Institut d'Astrophysique de Paris, Paris France\\
Institut d'Astrophysique Spatiale, Orsay France\\
Institut des Sciences Nucléaires de Grenoble, Grenoble France\\
IROE CNR, Firenze Italy\\
Jet Propulsion Laboratory, Pasadena USA\\
Laboratoire d'Astrophysique de l'Observatoire de Grenoble, Grenoble France\\
Laboratoire de l'accélérateur linéaire, Orsay France\\
Landau Institute of Theoretical Physics, Moscow Russia\\
Observatoire Midi-Pyrénées, Toulouse France\\
Queen Mary and Westfield College, London UK\\
Universita di Roma La Sapienza, Roma Italy\\
University of Minnesota at Minneapolis USA\\
}
}
\fi
\copyrightyear  {2001}

\begin{abstract}
  The Cosmic Microwave Background Radiation is the oldest photon
  radiation that can be observed, having been emitted when the
  Universe was about 300,000 year old. It is a blackbody at 2.73~K,
  and is almost perfectly isotropic, the anisotropies being about one
  part to 100,000. However, these anisotropies, detected by the COBE
  satellite in 1992, constrain the cosmological parameters such as the
  curvature of the Universe.
  
  Archeops is a balloon-borne experiment designed to map these
  anisotropies. The instrument is composed of a 1.5 m telescope and
  bolometers cooled at 85~mK to detect radiation between 150 and
  550~GHz. To lower atmosphere parasitic signal, the instrument is
  lifted at 32~km altitude with a stratospheric balloon during the
  arctic night.  This instrument is also a preparation for the Planck
  satellite mission, as its design is similar.

We discuss here the results of the first scientific flight from
Esrange (near Kiruna, Sweden) to Russia on January 29th 2001, which
led to a 22\% (sub)millimetre sky coverage unprecedented at this resolution.

\end{abstract}

\date{\today}

\maketitle

\section{The scientific objective}

\subsection{The Cosmic Microwave Background}

The Cosmic Microwave Background Radiation (CMBR) was emitted by the
Universe when it was 300,000 years old just after the Big Bang. Its
spectrum is known as a blackbody with a temperature of only
2.725~degrees above absolute zero. In various directions in the sky,
we observe small temperature differences of the order of one part in
100000, that were measured for the first time by the COBE satellite
\cite{Smoot}. These so-called anisotropies trace the 
fluctuations of the density of matter that occured before the
decoupling of the CMBR.  These fluctuations are thought to be the
origin, by gravitational collapse, of the large-scale structure of the
Universe (galaxies, clusters,...) that we observe today. Its pattern
can also yield an indirect measurement of the density, age and
curvature of the Universe (see \eg\cite{Hu}).

There have been many experiments that have already measured these
anisotropies with various techniques, angular resolution, noise and
scanning strategy. Most recent ones (e.g. TOCO,
Boomerang~\cite{deBernardis,Netterfield}, and
Maxima~\cite{Hanany,Lee}) have improved on COBE results by the
wavelength coverage, the sensitivity and the angular resolution.

\subsection{The observation strategy}

Balloon experiments are either limited by integration time due to
small duration flights (in USA or Europe) or Sun disturbance (in
Antarctic Summer). This in turn forbids mapping large portions of the
sky. An alternative is to use a flight during the polar night in the
more accessible Arctic region.

The Archeops experiment\footnote{More details on the experiment can be
  found at http://www.archeops.org} aims at mapping the anisotropies
of the cosmic microwave background from small to large scales at the
same time. For this purpose, a beam of about 8 arcminutes is swept
through the sky by spinning a 1.5 m telescope pointing at 41~degree
elevation around its vertical axis. A large fraction of the sky is
covered when the rotation of the Earth makes the swept circle drift
across the celestial sphere. This is only possible if the observations
are done during the Arctic night and on a balloon where neither the
Sun nor the atmosphere disturb the measurements. Ozone cloud emission
and residual winds can be avoided with a high altitude strastospheric
balloon.

From the Swedish balloon and rocket base in Esrange near Kiruna, in
cooperation with Russian scientists, the CNES balloon team can launch
balloons in the polar night, with a typical trajectory ending just
before the Ural mountains in Russia. Integration times can be up to
24~hours in the December-January campaigns.

\section{The instrument}

A general description of the first Archeops instrument can be found in
\cite{Benoit} where the first gondola used during the test flight
(that happened in Trapani in July 1999) is described. The present
experiment mainly uses the same concept.

\subsection{The telescope, optics and detectors}

The Archeops telescope is a two mirror, off-axis, tilted Gregorian
telescope consisting of a parabolic primary (main diameter of 1.5~m
diameter) and an elliptical secondary (this design is similar to the
one proposed for Planck during phase-A). The telescope was designed to
provide diffraction-limited performance when coupled to single mode
horns producing beams with FWHM of 8~arcminutes or less at frequencies
higher than 140~GHz. Both mirrors were milled from 8 inch thick
billets of aluminum 6061-T6 and were thermally cycled twice during
machining to relieve internal stresses. The primary and secondary
mirrors weigh 45~kg and 10~kg respectively.

\begin{figure}[htbp]
  \includegraphics[angle=90,height=.3\textheight]
    {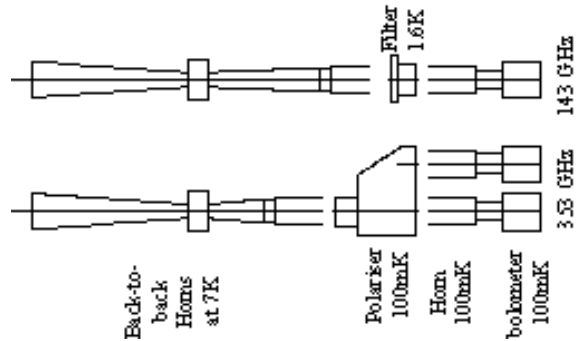}
      \caption{The cold optics (of which the 143~GHz channel is shown
        on the top) is made of a 10~K back-to-back horn, a 1.6~K
        filter stage and the 100~mK horn and bolometer stage. Light
        enters from the left. A typical 353~GHz channel has one 10~K
        stage for two bolometers.  A polariser splitter is inserted at
        the 1.6~K stage (bottom configuration). }
         \label{fig:photpix}
   \end{figure}

For CMB anisotropy measurements, control of spectral leaks and beam
sidelobe response is critical. Archeops channels have been
specifically designed to maximize the sensitivity to the desired
signal, while rejecting out-of-band or out-of-beam radiation. We have
chosen to use the configuration developed for Planck HFI, using a
triple horn configuration for each photometric pixel, as shown
schematically in Fig.~\ref{fig:photpix}.

In this scheme, radiation from the telescope is focussed into the
entrance of a back-to-back horn pair. With no optical components in
the path, control of the beam is close to ideal. Proper single mode
corrugated feeds associated with a new profiled-flared design will be
used to obtain 30dB telescope edge taper with the telescope/horns
combination. The new profiled-flared horns avoid the use of a lens at
the exit aperture of the second horn, creating a beam-waist where
wavelength selective filters can be placed. Finally, the third horn
maintains beam control and focuses the radiation onto the spider web
bolometer placed at the exit aperture.

A convenient aspect of this arrangement is that the various components
can be placed on different temperature stages in order to create
thermal breaks and to reduce the level of background power falling
onto the bolometer and fridge. In Archeops, the back-to-back horn pair
is located on a cold plate cooled by Helium vapor at 8~K. Sidelobe
response, beamwidth on the sky and spillover are accurately controlled
by the design of the front horn. More low pass filters are placed on a
screen at 1.6~K to further reject unwanted radiation from the inner
sanctum where the 100~mK detectors are located.

Twenty two bolometers are placed on the 100~mK low temperature plate.
There are 9 bolometers at 143~GHz, 7 at 217~GHz , 6 polarised
bolometers at 353~GHz and two at 545~GHz. The higher frequency horns
(545 GHz) are multimoded, as this increases the signal at this
frequency and the side lobes rejection is less critical. One blind
bolometer is placed on the same copper plate to study the electronic
noise of the bolometers at this stage. These are placed at different
points in the focal plane and observe the same sky pixel at a
different time. Bolometers on the same line observe the sky with
typicaly 100~msec time difference as bolometers on different lines
observe the same pixel with a time difference of the order of a few
minutes.  The six 353~GHz channels are devoted to the measurement of
galactic polarized emission. The bolometers are assembled in three
pairs, with one single back-to-back horn and a polarizer splitter. The
two bolometers of each pair measure the polarized intensity of the
incoming signal in two orthogonal directions. Each pair makes a
different angle with respect to the scan axis to enable the full
determination of the Stokes parameters. Archeops will provide the
first measurement of polarization in this range of frequencies with a
sensitivity adequate for measuring galactic dust polarised emission,
as well as a validation of the technical configuration for PLANCK-HFI.

\subsection{The gondola and the pivot}
\label{sec:gondola}

The gondola is made with welded aluminium square tubes 30*30*2 and a
careful design prevents from important deformations of the optical
design in the presence of strength. Typical change in the relative
mirror position stays below 0.2~mm when the gondola is lifted or
tilted. Total mass of the main frame is 70~kg. The two mirrors and the
cryostat are fixed to the frame and the elevation of the beam
direction is fixed to the value of 41~degrees.

The pivot connects the flight chain of the balloon to the payload
through a thrust bearing, providing the necessary degree of freedom
for payload spin. Two deep groove bearings provide stiffness against
transverse loads to the rotating steel shaft inside the pivot. The
pivot includes a torque motor that acts against the flight chain to
spin the payload. After initial acceleration, the motor provides just
enough torque to compensate the friction in the thrust bearing and the
small residual air friction.

The rotation of the payload is monitored by a vibrating structure rate
gyroscopes that can detect angular speeds as low as 0.1~deg/s. These
are sampled at 150~Hz by a 16~bit ADC and a PID feedback loop control
is implemented in software, to drive the torque motor in the pivot.

\subsection{The fast stellar sensor}
\label{sec:FSS}

A custom star sensor has been developed for pointing reconstruction in
order to be fast enough to work on a payload rotating at 2-3~rpm. At
this spin rate, the use of a pointed platform for the star sensor is
impractical. Each independent beam (8~arcmin. wide) is scanned by the
mm-wave telescope in about 10~ms, establishing a detector response
time that excludes the use of present large-format CCDs.

We decided therefore to develop a simple night sensor, based on a
telescope with photodiodes along the boresight of the mm-wave
telescope. Thus, like the millimeter telescope, the star sensor scans
the sky along a circle at an elevation of 41~deg.

A linear array of 46 sensitive photodiodes (Hamamatsu S-4111-46Q) were
placed in the focal plane of a 40~cm diameter, 1.8~m focal length
parabolic optical mirror. Each photodiode has a sensitive area of 4~mm
(in the scan direction) by 1~mm (pitch in the cross-scan direction).
The line of photodiodes is perpendicular to the scan and covers
1.4~degrees in elevation on the sky, with about 7.6~arcminute (along
the scan) by 1.9~arcminutes (cross-scan) per photodiode. A top baffle,
painted black inside and located above all nearby payload structures,
prevents stray radiation.

The sensitivity of the photodiodes defines the average number of stars
we can observe during one rotation of the payload. With a sensitivity
limited to stars of magnitude 7, we can count between 50 and 100 stars
per turn during night time. In order to control the pointing during
the day, we use an optical filter in front of the diodes to minimize
the perturbation due to stray light. A test flight in Kiruna (April
1999) shows that, even with the Sun at low elevation (< 5~deg.
elevation) we can observe a few stars for each rotation of the
gondola.

The star sensor software extracts from the time-sampled photodiode
signals candidates star with detection time, measured flux, coordinate
along diode array, and quality criteria. This software produces from
raw data the list of time-ordered star candidates and makes it
available for the second step of the software-reconstruction of the
telescope pointing. The attitude reconstruction algorithm is based
upon the comparison between star candidates and a dedicated star
catalog. The fluxes of stars in catalog are computed from the
Hipparcos catalog to simulate the star sensor spectral response.

The reconstruction is achieved using only star sensor data if the gondola
spin axis motion is sufficiently slow. That was the case for the
Trapani test flight. The precision of the pointing solution is better
than 1~arcminute rms for the test flight. For the Kiruna 2001 flight,
important speed variation of gondola rotation velocity required to use
additional information from gyroscopes and gps to recover a good
association between signal and star catalog.

\subsection{The cryogenics}
\label{sec:cryo}

The focal plane is cooled to 100~mK by means of an open cycle dilution
refrigerator. This type of refrigerator has been designed for
satellite applications (it will be used on Planck HFI) and Archeops is
the first balloon-borne experiment using a dilution refrigerator. The
dilution stage is placed in a low temperature box placed on the top of
a liquid Helium reservoir at 4.2~K. The top part of this box contains
the entrance horns and receives a significant amount of heat power
from near infrared radiation (about 500~mW). Exhaust vapours from the
helium tank maintain the horns near 7~K. The entrance is
protected from radiation by two vapour cooled screens with openings
for the input beam. The filters are placed on the horns at 7~K, on the
1.6~K stage (cooled by Joule-Thompson expansion of the dilution
mixture) and on the 100~mK stage, just in front of the bolometers. The
temperatures of each stage are monitored with thermometers: carbon
resistance and NbSi metal insulation transition thermometers.

The bolometers are placed on the 100~mK stage supported by Kevlar
cords. The dilution fluids (isotopic pure $^3$He and $^4$He) arrive through
two small capillary tubes along a heat exchanger with the return
mixture. The two capillaries join and the $^3$He is dissolved into the
$^4$He, cooling down the mixture which is used to cool down the 100~mK
plate using a small heat exchanger. We use two extra capillaries of
larger diameter (0.5~mm) in order to precool the system by a
circulation of $^4$He gas. These 5 capillary tubes (3 for dilution and 2
for precooling) and the electric wires (9 shielded cables with 12
conductors each) are soldered together, forming the continuous heat
exchanger disposed around the 100~mK stage. Input flow is controlled
by an electronic flow regulator. The output mixture is pumped with a
charcoal pump placed inside the liquid helium (1 liter box filled with
charcoal). During pre-launch operations, the output mixture is
extracted with an external pump and the dilution stage can stay below
100~mK continuously for months. The hold-time of the cryostat is
limited at 48~hours by the liquid helium tank of 20~liters. An
electronic regulator is used to maintain constant pressure at one
atmosphere in the helium tank.

In order to insure temperature stability of the bolometer a passive
filter is used to thermalise the bolometer plate. The open cycle
dilution produces large temperature fluctuation ($100\, \mathrm{\mu K
  / sqrt(Hz)}$) which we attenuate with a high specific heat material
(HoY). Holmium has a Shottky anomaly around 200~mK which insures the
high specific heat. Mixing it with Yttrium helps controlling the
conductivity. By cooling down the bolometers through this thermal
filter, a stability below a few $\, \mathrm{\mu K / sqrt(Hz)}$ was
obtained during the flight.

\subsection{The electronics}
\label{sec:elec}

The bolometers are biased using AC square waves by a capacitive
current source. Their output is measured with a differential
preamplifier (the first stage uses JFET working at about 120~K) and
digitized before demodulation. We use the boxes already designed in
preparation for Planck HFI instrument. Each box can manage 6
bolometers and we used 6 of them for a total of 36 channels. All
modulations are synchronous and driven by the same clock. This clock
is also used for data readout, which is simultaneous for all
bolometers and thermometers. Modulation parameters can be controlled
by telecommand. Sampling of the raw signal is at 6.51~kHz before
demodulation. Demodulation is performed by the EPLD and sampled twice
per modulation period. We used a frequency of 76~Hz for modulation and
152~Hz for sampling. The on-board computer uses a transputer T805 and
an 4 EPLD Altera 9400 to control all houskeeping measurement,
bolometer and star sensor data. The power supply consists of 39
batteries for electronics and satellite telemetry and 36 for the
motor. All are 3~V and 36~Ah lithium batteries. With a total power of
150~W, this gives us about 48~h lifetime for the experiment.

\subsection{The telemetry}
\label{sec:tel}

After compression, the data are written to a storage module of 2~Gbyte
Flash Eprom memory made of 256 circuits of 8 Mbyte each. A dedicated
microprocessor is used to write the Eprom. To protect the data in case
of bad landing, the data storage module is installed in a sealed box,
pressurized at 1 atmosphere. The data are read after retrieval of the
gondola. The compressed data are sent via the standard CNES telemetry
(400~MHz) at the rate of 108~kbit/sec. This is possible only during
the first phase of the flight (about 4 hours) when the balloon is in
direct contact with the ground telemetry station. Another telemetry
channel using the Inmarsat satellite is used to control the experiment
during all the flight. We use the mini-M Inmarsat standard that allows
a typical flow rate of 2~kbit/sec. A selected fraction of the data is
sent through this channel to control the experiment and commands can
be sent to correct all control parameters.

\section{Archeops flights and first results}
\label{sec:flight}

A first flight of the instrument took place in Trapani on July 17th
1999. This test flight used only a few detectors (5) and we got only 4
hours of data during the night. Nevertheless, this flight allowed us
to check all the fonctionnalities of the instrument~\cite{Benoit}.

\subsection{Flight conditions at Esrange in 2000-2001 Winter}
\label{sec:conditions}

This Winter (December 2000 and January 2001), the polar vortex was not
well positioned as the 2 previous winters and we did not get good
flight conditions during the campaign. The wind conditions in the
stratosphere gave us the possibility of a long flight only on December
1rst, where the Archeops technical flight took place. Later on, we did
not get good conditions before January 12th 2001 where we launched
Archeops for a 7-8h flight. However, one of the flow-meter controlling
the cryostat broke down 1~hour after take off and we had to abort the
flight with a landing in Finland and a fast recovery without too much
damage on the instrument. The next launch window opened on January
29th for a relatively short duration flight at an altitude of 32~km
(too much wind at higher altitude) lower than the nominal 40~km.

\subsection{The scientific flight}
\label{sec:KS1}

We finally were able to launch Archeops for its first scientific
flight on the 29th of January, 2001. Because of the wind conditions we
could only use a smaller balloon (150 000 m$^3$) otherwise the
trajectory would have been too much North, endangering the recovery.

We had 7h30 of flight at float, with a temperature of the order of
90~mK on the focal plane almost all the way along. We decided not to
fly too close to the Ural mountains: this would have been too risky
for the recovery because of strong wind. The experiment was stopped
(window closed and motor stopped) about one minute before the
separation with the balloon occured thanks to Inmarsat (00h26m30 LT).
The gondola reached the ground at a latitude of 62.226 and a longitude
of 53.341 (N-E of SyktYvkar).

\subsection{Preliminary results}
\label{sec:prelim}

The first task after the flight is to reconstruct the pointing using
the data from the stellar sensor. The residual error in pointing is
given by the distribution of the declination difference between the
reconstructed position and the stars. The present precision of
the pointing solution for this flight is better than 2 arcminute rms.

To control the accuracy of the reconstructed pointing and extract the
in-flight angular resolution of the detectors, their time response,
and to calibrate the instrument on point sources, we use the signal of
Jupiter measured in the bolometers and the pointing of each detector
reconstructed from the Stellar Sensor. Jupiter is a bright source
which is seen by one bolometer at a time thanks to Archeops scan
strategy, we can therefore easily pinpoint its crossings (we could
see it twice during the January 2001 flight).

\begin{figure}[htbp]
  \includegraphics[angle=0,height=.3\textheight]
    {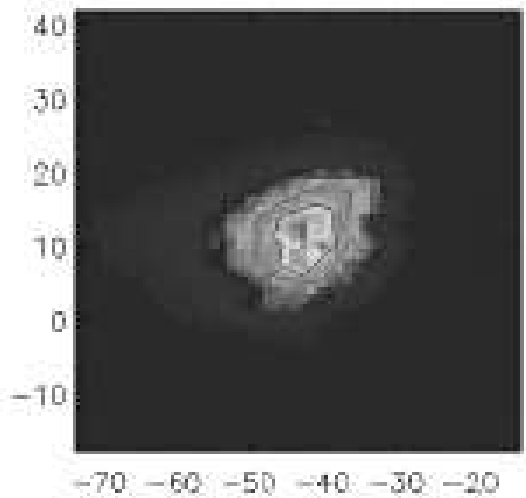}
      \caption{Jupiter as observed by one of the bolometers at 217~GHz. 
Azimuth-Elevation coordinates are in arcminutes. Note that the
bolometer time constant spreads the signal a little on the left.}
         \label{fig:jup}
   \end{figure}

If the pointing is accurate enough on the corresponding time periods,
we can reconstruct the angular resolution as shown on
Fig.~\ref{fig:jup}, where the beam of one of the 217~GHz detector is
represented as obtained on Jupiter: we found the FWHM of this
particular beam to be of the order of 12.5 arcminutes, which includes
the bolometer and electronics time response: the deduced optical beam
is nominal at 8~arcminutes.

Since the response of the bolometers to Jupiter's signal is a
convolution of the intrinsic resolution of the back-to-back horns and
the time response of the bolometer, we can also extract a measurement
of both parameters: we typically found a time constant of the order of
5 to 15~ms, compatible with measurements on glitches (the intrinsic
resolution of the horns being between 5 and 8 arcminutes).

As far as the calibration on point sources is concerned, we make use
of the temperature of Jupiter~\cite{Goldin} (T = 170 K). Taking into
account its solid angle, we have extracted a calibration for each
Archeops bolometers.

\begin{figure}[htbp]
  \includegraphics[angle=0,height=.3\textheight]
    {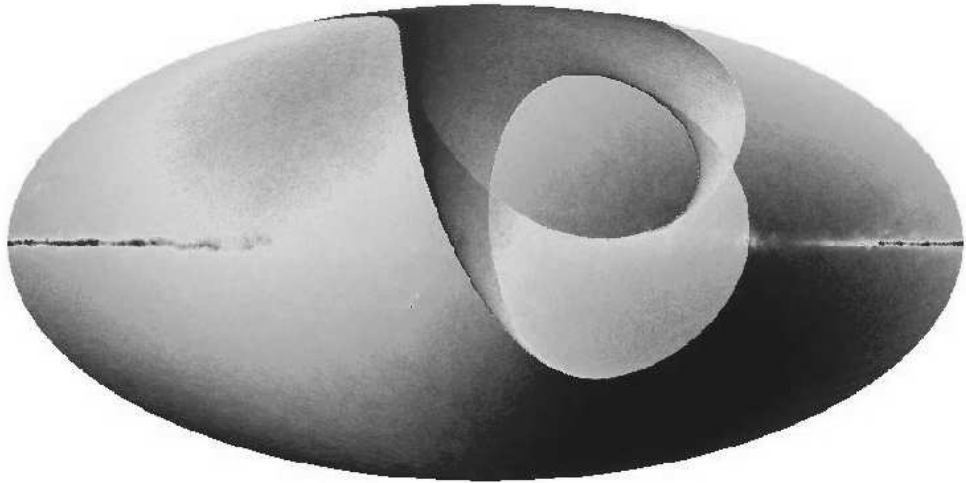}
      \caption{Expected 143~GHz map with the CMB dipole, anisotropies
        and IRAS/DIRBE extrapolation of galactic dust emission
        (\cite{Schlegel}). This is an all-sky Mollweide projection
        centered on the Galactic anticenter. Overimpression shows the
        area covered during last Kiruna flight. Revolutions started in
        the lower right and finished in the upper left. Most of the
        sky above the galactic plane can be used for CMB anisotropy
        measurements }
         \label{fig:moll}
   \end{figure}

   The 143 and 217~GHz channels are dominated by the cosmic dipole and
   some extra signal coming from the 10~K back-to-back horn emission
   (sinusoidal shape). At 545~GHz, the emission from the Galaxy is
   dominant as well as some atmospheric signal.
   
   Flight at ceiling represents 7.5~hours worth of scientific data
   taken when the gondola was spinning at 2~rpm. The covered area
   corresponds to 22~percent of the whole sky. Fig.~\ref{fig:moll}
   shows a Mollweide projection in galactic coordinates (centered on
   the Galactic anticenter) of an extrapolation to the submillimetre
   of a combined IRAS-DIRBE map~\cite{Schlegel}.  Due to the relative
   small duration (compared to 24h of one earth rotation), the area
   where the circles cross each other is very small. As a consequence,
   we had a relatively poor redundancy during this flight.
   
   The galactic plane is well observed at all frequencies from Perseus
   to Cygnus regions. Some clouds much below the Galactic plane can
   easily be identified with their CO and infrared counterparts
   (Taurus, Pleiades, ...).

   With the 353~GHz channels, Archeops will provide the first
   measurement of galactic polarized emission in this range of
   frequencies. It is first an important topic in the prospect of
   foreground removal for Planck-HFI, and is also of great interest to
   constrain the physics of galactic dust and molecular clouds.
   
\begin{figure}[htbp]
  \includegraphics[angle=0,height=.4\textheight]
    {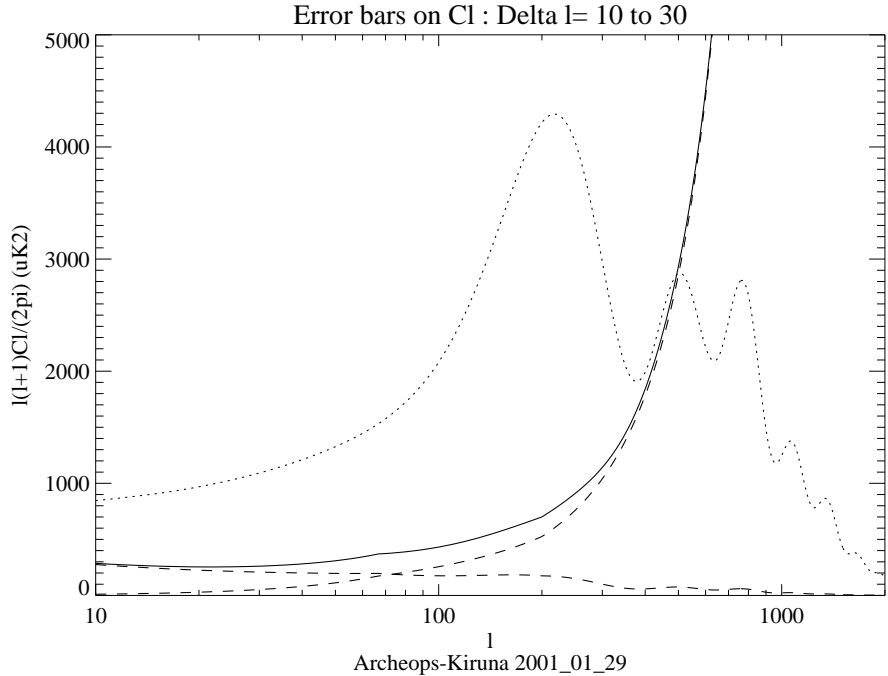}
      \caption{Sensitivity expected for the last Kiruna flight. The
        dotted curve shows a fiducial $C_l$ spectrum for a standard
        cosmology (in $\mu K^2$). The continuous line is the expected
        $1\sigma$ error using the characteristics of the instrument
        during the flight (flat noise average and responsivity). The
        first dashed curve (on the left) shows the cosmic variance
        part of the noise and the second (on the right) shows the
        effect of detector noise. }
         \label{fig:cl}
   \end{figure}

   Sensitivities are typically between 50 and $100\zu \mu K_{RJ}$ for
   one second of integration for one photometric pixel. There are
   about 8 pixels with a CMB sensitivity between 120 and $200 \zu \mu
   K_{CMB}$ for one second of integration for one photometric pixel.
   This instrument is therefore very competitive with respect to other
   designs. Work is in progress to subtract all parasitic signals and
   extract the CMB fluctuation spectrum. Expected performances for the
   present flight are shown in Fig.~\ref{fig:cl}. Good detections of
   the CMB anisotropy spectrum can be expected from low l to beyond
   the first acoustic peak.

\section{Conclusions and future flights}
\label{sec:conc}

Archeops is a balloon borne experiment dedicated to the measurement of
the anisotropies of the Cosmic Microwave Background using the same
technology as the Planck satellite will use. We were able to launch
the instrument on the 29th of January 2001 at Esrange (Sweden). The
flight lasted 7h30 at float, with all detectors working with nominal
performances. This is a very important step in validating Planck-HFI
concepts (Lamarre \etal, this conference). The flight was unusually
short due to strong winds in the stratosphere. We are currently
analysing the data: reconstructing the pointing using the Stellar
Sensor data, calibrating the instrument using point sources (Jupiter)
as well as the dipole and the galaxy, and already some results are
coming along such as maps of the galactic plane.

Nevertheless, because of the strong winds, the movements of the
gondola induced a parasitic signal on bolometers which is different
for each frequency. This makes the data analysis more difficult than
planned. A flight at higher altitude (with a 400000 or 600000 m3
balloon) should permit to have a more stable experiment, with less
atmospheric signal.  In addition, to have a better measurement of the
Cosmic Microwave Background, a longer flight (a 24 hour flight for
instance or two flights with 12 hours each) should provide a larger
redundancy on the sky. Expected sensitivity shows that significant
detections of the $C_l$ spectrum could be achieved between $l=10$ to
1000. The large area covered by Archeops makes it a unique balloon
instrument in lowering the CMB cosmic variance, the most important
noise at $l$ below 100 (see Fig.~\ref{fig:cl}). A new campaign is
planned for the Winter 2001/2002, with two launch windows: one in
December and one in January.

\begin{theacknowledgments}
  We thank the CNES and Esrange Swedish Facility for their continued
  support for this project and the flights (technical and scientific)
  that were realised very smoothly.
\end{theacknowledgments}

\end{document}